\documentclass[mathleft
]{an}
\usepackage{graphicx}
\usepackage{times}
\begin{document}

\Pagespan{789}{}
\Yearpublication{2011}%
\Yearsubmission{2011}%
\Month{11}%
\Volume{999}%
\Issue{88}%

\title{A new flare star member candidate in the Pleiades cluster}

\author{M. Moualla\inst{1}\fnmsep\thanks{Corresponding author:
  \email{mmoualla@astro.uni-jena.de}\newline}, {T. O. B. Schmidt\inst{1}}, {R. Neuh\"auser\inst{1}}, 
{V. V.  Hambaryan\inst{1}}, {R. Errmann\inst{1}}, {L. Trepl\inst{1}} \and Ch. Broeg\inst{2}, 
T. Eisenbeiss\inst{1}, M. Mugrauer\inst{1},
C. Marka\inst{1}, C. Adam\inst{1}, C. Ginski\inst{1}, T. Pribulla\inst{1,3}, 
S. R\"atz\inst{1}, J. Schmidt\inst{1}, A. Berndt\inst{1}, 
G. Maciejewski\inst{1,4}, 
T. R\"oll\inst{1}, M. M. Hohle\inst{1}, N. Tetzlaff\inst{1},
S. Fiedler\inst{1} 
\and S. Baar\inst{1}
}
\titlerunning{A new flaring member candidate in the Pleiades cluster}
\authorrunning{Moualla et al.}
\institute{
Astrophysikalisches Institut und Universit\"ats-Sternwarte Jena, FSU Jena, Schillerg\"a\ss chen 2-3, 07745 Jena, Germany
\and 
Space Research and Planetary Sciences, Physikalisches Institut, University of Bern, Sidlerstra\ss e 5, 3012 Bern, Switzerland
\and
Astronomical Institute, Slovak Academy of Sciences, 059 60, Tatransk\'a Lomnica, Slovakia
\and
Toru\'n Centre for Astronomy, Nicolaus Copernicus University, Gagarina 11, PL–87-100 Toru\'n, Poland
}

\received{2011}
\accepted{}
\publonline{later}

\keywords{open clusters and associations: individual (Pleiades) -- stars: low-mass, late-type, flare -- techniques: photometric}

\abstract{We present a new flare star, which was discovered during our survey on a selected field at the edge of the Pleiades cluster. The field was observed in the period 2007 -- 2010 with three different CCD-cameras at the University Observatory Jena with telescopes from 25 to 90\,cm. The flare duration is almost one hour with an amplitude in the R-band of about 1.08\, mag. The location of the flare star in a color-magnitude diagram is consistent with age and distance of the Pleiades. In the optical PSF of the flare star there are two 2MASS objects (unresolved in most images in the optical Jena PSF), so it is not yet known which one of them is responsible for this flare. The BVRIJHK colors yield spectral types of M1 and M2 with extinction being $A_{V}$\,=\,0.231\,$\pm$\,0.024\,mag and $A_{V}$\,=\,0.266\,$\pm$\,0.020 for those two stars, consistent with the Pleiades cluster.}

\maketitle

\section{Introduction}

The first flare star was discovered in 1924 by Hertzsprung (Hertzsprung, 1924). Later a new type of flares was discovered in the solar neighborhood and carried the name UV Cet-stars, which are known to be low-mass stars with very low luminosities. The investigation and study of the flares in nearby open clusters such as the Pleiades showed their importance for the evolution of young low-mass stars and protostars (Mirzoyan et al. 1995). The first flare star in the Pleiades was discovered by Johnson \& Mitchel (1958). By studying the statistical characteristics of flare stars in the Pleiades by Ambartsumyan et al. (1970, 1971), it was found that almost all stars fainter than $V$\,=\,13.3\,mag can be flare stars. Another advantage of flare star investigations in nearby clusters is to identify additional cluster members at very low masses. Most of the Pleiades faint members were discovered in flare star surveys (Jones 1981; Haro et al. 1982; Stauffer et al. 1991).

A field at the edge of the Pleiades cluster was chosen to be observed in our survey to find new low-mass stars or brown dwarfs and variable stars, because in this field there was no survey for brown dwarfs done before (see Table 1 in Schwarz \& Becklin 2005). One new flare star was found, which lies on the Pleiades zero-age main-sequence, and is presented in this paper. In Sect. 2 we describe our observations and data reduction. The photometric analysis of our data is shown in Sec. 3, results are presented and discussed in Sect. 4.   

\section{Observations and data reduction}

Our field was chosen to be at the edge of the Pleiades cluster, in order to look for 
new brown dwarfs and low-mass stars by deep imaging and follow-up spectroscopy, 
since no deep CCD surveys were done there before; at the same time, 
we wanted to study the variability of all stars in the field on 
timescales of minutes to years. Concerning brown dwarfs several candidates were found in 
our field (Eisenbeiss et al. 2009; Seeliger et al. 2011, in preparation).

We observed two overlapping sub-fields with field centers at
$\alpha _{J2000}$ = $\mathrm{3^h}$ $\mathrm{42^m}$ $\mathrm{20.6^s}$, $\delta _{J2000}$ = +25$^\circ$ 36$'$ 54$''$
and also at
$\alpha _{J2000}$ = $\mathrm{3^h}$ $\mathrm{40^m}$ $\mathrm{54^s}$, $\delta _{J2000}$ = +25$^\circ$ 40$'$ 32$''$.

All observations were carried out at the University Observatory Jena, which is located close to the village Gro\ss schwabhausen (GSH), about 10 km to the west of Jena. In March 2007 we started our survey using the 0.25\,m Cassegrain-Teleskop-Kamera (CTK), which is installed at the 0.25\,m auxiliary Cassegrain telescope, mounted at the tube of the 0.9\,m telescope (see Mugrauer 2009). It has a field of view (FoV) of 37.7$'$ $\times$ 37.7$'$. In August 2010 the CTK was replaced with a new CCD-camera called CTK-II, which has a FoV of 21.0$'$ $\times$ 20.4$'$ (Mugrauer 2011, in preparation). Since beginning of February 2009 the new Schmidt-Teleskop-Kamera (STK) is in operation in the Schmidt-focus of the 0.9\,m telescope (0.6\,m in Schmidt mode) at the University Observatory Jena. With its 2048$\times$2048 pixels CCD detector and a pixel-scale of about 1.5\,arcsec/pixel it has a FoV
of 52.8$'$ $\times$ 52.8$'$ (see Mugrauer \& Berthold 2010). Since the STK camera was used mainly in our survey due to its larger FoV in comparison to the other cameras we list in Table \ref{table2} only the observations with the STK camera.
Observations were done mostly in R-band. At the beginning of our survey with the CTK we used an individual exposure time of 60\,s per image. The same exposure time was used later with the STK. We found many new variable stars in the field (Moualla 2011). One of them was very faint, so that we increased our exposure time to 90 s.
Basic data reduction was done for all images, by means of dark subtraction and flat fielding using the ESO software MIDAS (Munich Image Data Analysis System).

\section{Photometry}

After reducing the images, three different programs were used to obtain 
positions and magnitudes of all objects in the field.\\
First the source extractor (SE) for the source detection with GAIA (Graphical Astronomy 
and Image Analysis Tool) was performed on each single frame in order to distinguish 
between real and false objects by performing thresholding and deblending (see Bertin 1997) as 
well as to determine their center coordinates ($\mathrm{X_{Cen}}$, $\mathrm{Y_{Cen}}$).\\
SE uses automatic aperture photometry, originally designed for galaxies and extended 
objects; especially in crowded fields as well as undersampled point sources, this method fails 
to obtain accurate magnitudes; furthermore, their errors are underestimated; however, source detection 
based on thresholding, as SE employs, is a fast and convenient way to detect even faint objects 
with reasonable positional accuracy.\\
The instrumental magnitudes for all objects detected by SE were determined then
by using aperture photometry with MIDAS. We created a program for the aperture
photometry with MIDAS which needs the center coordinates of all objects in the field.
The flux of the star was measured with an internal aperture with a radius of 5 pixels
and the background was measured 20 pixels away around the internal one.\\
The differential magnitudes of all objects in the field were determined by using an algorithm (Broeg et al. 2005) that uses all stars in the field as comparison stars in order to determine the best synthetic comparison star by calculating their weighted average. Very small weights were given to unsuitable stars that are possibly variable or very faint. By excluding these unsuitable stars the best synthetic comparison star is calculated and the differential magnitudes of all objects in the field are determined by comparing their instrumental magnitudes to the magnitude of the synthetic comparison star.

\begin{figure*}                                                                                                                                                \centering                                                                                                                                                     \includegraphics[width=12cm]{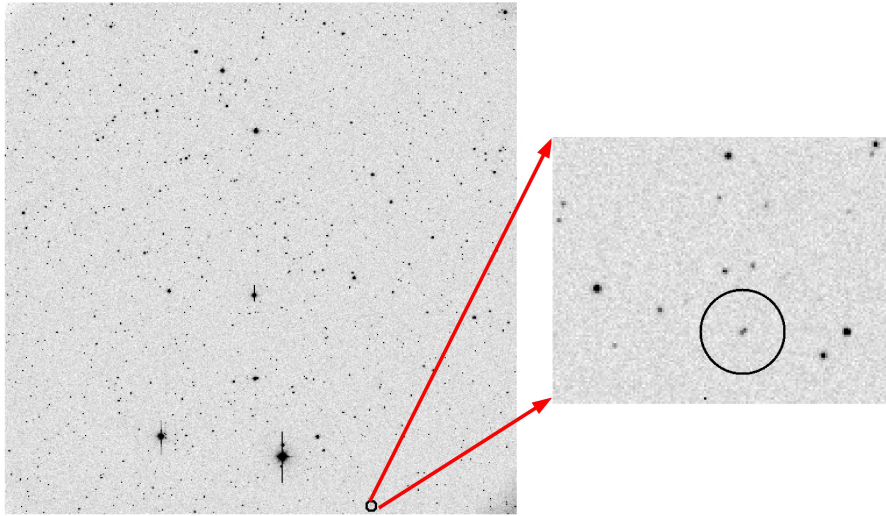}                                                                                                                      \caption{The left panel shows the Pleiades field with the flare star observed by the STK camera in R--band in the night 2010 September 20 with an exposure time of 90s. The field center on the left is $\alpha _{J2000}$ = $\mathrm{3^h}$ $\mathrm{40^m}$ $\mathrm{54^s}$, $\delta _{J2000}$ = +25$^\circ$ 40$'$ 32$''$.
The right panel zooms into the FoV of the STK, the circle marks the new flare star. The FoV of the left image is 52.8$'$ $\times$ 52.8$'$ and the right field is 3.9$'$ $\times$ 2.9$'$. North is up, East is to the left.}                                                                                                              \label{Flare1}                                                                                                                                                 \end{figure*}                                                                                                                                                  
\begin{figure}                                                                                                                                                 \centering                                                                                                                                                     \includegraphics[width=6cm]{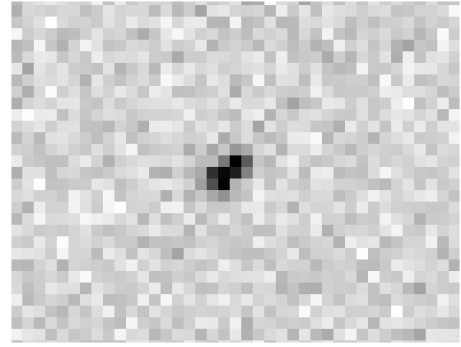}                                                                                                                        \caption{A zoom (1$'$ $\times$ 1.2$'$) into the FoV of the STK camera shows that the new flare star, observed in R--band with an exposure time of 90s, is one of two stars (unresolved in most images, so that individual light curves cannot be obtained). North is up, East is to the left.}                          \label{flare2}                                                                                                                                                 \end{figure}                                                                                                                                                   
\begin{figure}                                                                                                                                                 \centering                                                                                                                                                     \includegraphics[width=6cm]{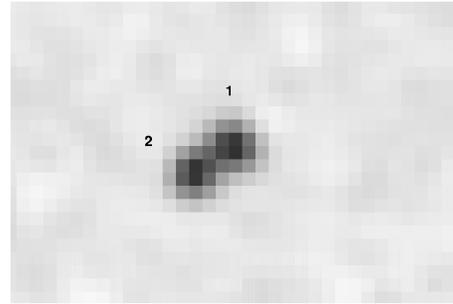}                                                                                                                         \caption{A zoom (33.80$''$ $\times$ 22.62$''$) into the Ks-band 2MASS image of our flare star (Skrutskie et al. 2006). North is up, East is to the left. Stars 1 and 2 as in Table 5.}                                                                                                                                             \label{2MASS}                                                                                                                                                  \end{figure}

\begin{table*}
\caption{List of constant stars in the field used to determine the zero point of the detector in VRI bands. In fall 2010 the V-filter showed impurity, so that the quality of the measurements in V is lower than in R and I.}
\label{table1}
\begin{tabular}{lcccccc}\hline
Star name&RA(J2000.0)&Dec(J2000.0)&($\mathrm{C_{V}}$\,-\,k$\cdot$z)&($\mathrm{C_{R}}$\,-\,k$\cdot$z)&($\mathrm{C_{I}}$\,-\,k$\cdot$z)&Spty\\
         &[h m s]&[$^\circ$ $'$ $''$]&&[mag]&&\\
\hline
Melotte 22 SK 785&03 40 14.80&+25 19 20.0&13.862\,$\pm$\,0.174&13.533\,$\pm$\,0.022&13.024\,$\pm$\,0.017&M3V\\
Melotte 22 MT 22&03 40 58.40&+25 17 43.0&13.357\,$\pm$\,0.268&13.297\,$\pm$\,0.016&12.402\,$\pm$\,0.013&A7V\\
Melotte 22 SK 792&03 40 05.00&+25 31 38.0&13.960\,$\pm$\,0.216&13.083\,$\pm$\,0.019&12.671\,$\pm$\,0.009&K4V\\
NVSS J034035+253859&03 40 35.57&+25 39 00.6&13.347\,$\pm$\,0.180&13.608\,$\pm$\,0.015&12.365\,$\pm$\,0.018&A8V\\
Melotte 22 DH 153&03 40 49.89&+26 03 33.5&13.935\,$\pm$\,0.191&13.387\,$\pm$\,0.014&12.626\,$\pm$\,0.009&M2V\\
Melotte 22 SK 769&03 40 24.60&+26 03 26.0&13.302\,$\pm$\,0.262&13.121\,$\pm$\,0.018&12.591\,$\pm$\,0.014&K4V\\
\hline
Mean value&&&13.627\,$\pm$\,0.215&13.338\,$\pm$\,0.0171&12.613\,$\pm$\,0.013&\\
\hline  
\end{tabular}                  
\end{table*}

\begin{figure*}
\centering
\includegraphics[width=12cm,angle=-90]{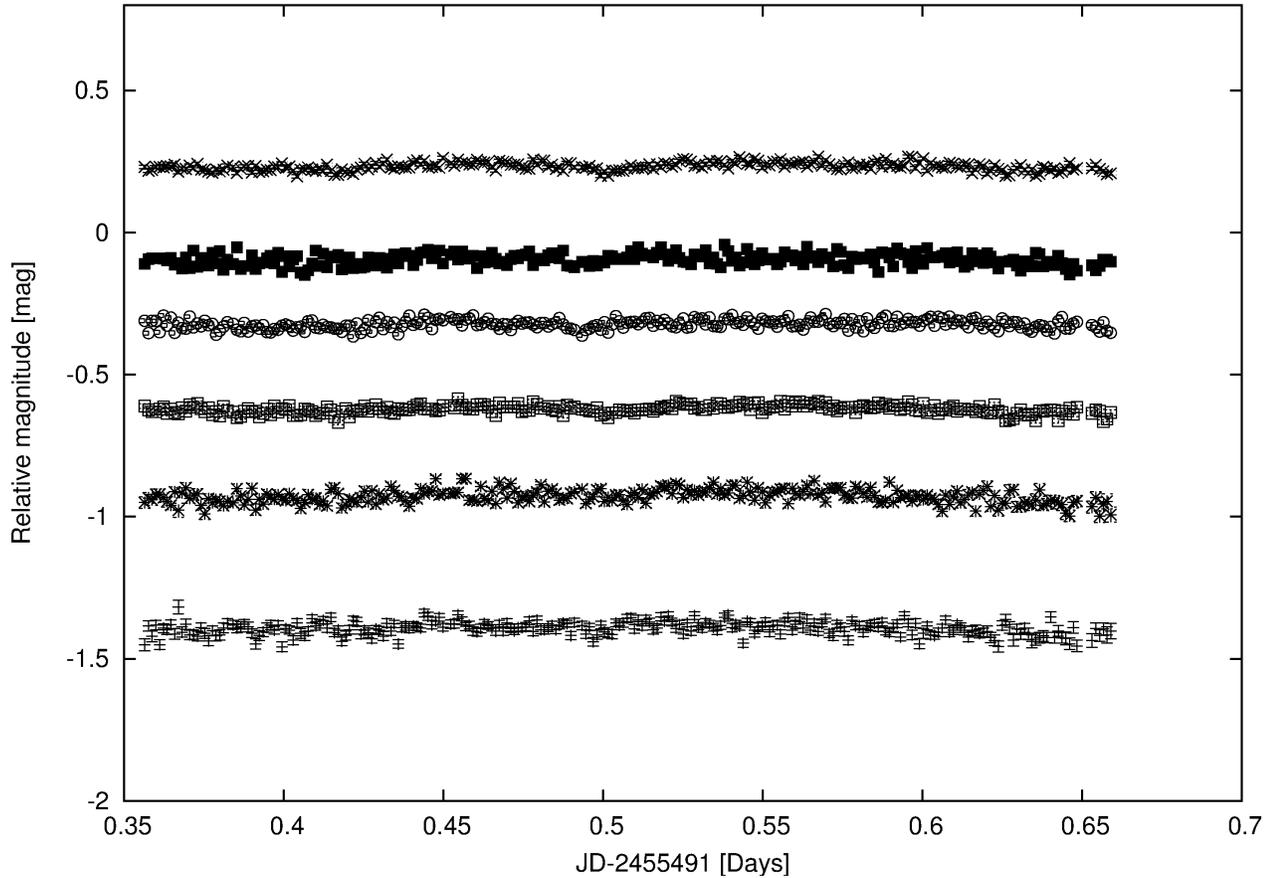}
\caption{Light curves of the constant stars in the field used to determine the zero point of the detector in R-band. They were observed in the night 2010 October 21 with the STK in R-band using an exposure time of 90\,s. Shown from top to bottom are Melotte 22 SK 785 with $\sigma$ = 0.0163\,mag, Mellote 22 MT 22 with $\sigma$ = 0.0057\,mag, Mellote 22 SK 792 with $\sigma$ = 0.0165\,mag, NVSS J034035+253859 with $\sigma$ = 0.0073\,mag, Mellote 22 DH 153 with $\sigma$ = 0.0065\,mag and Melotte 22 SK 769 with $\sigma$ = 0.0064\,mag (see Table \ref{table1}). }
\label{konstat} 
\end{figure*}

\begin{table}
\caption{Observations log of the Pleiades field with the STK.}
\label{table2}
\begin{tabular}{ccccc}\hline
Date&Interval&Filter&Expo.&N\\
&UT&Bessel&[s]&images\\
\hline
2009 Sep 18&01:50\,--\,03:43&R&60&96\\
2009 Sep 19&00:16\,--\,03:45&R&60&175\\
2009 Sep 21&22:41\,--\,01:51&R&60&161\\
2009 Sep 24&01:01\,--\,03:51&R&60&142\\
2009 Sep 26&02:56\,--\,03:59&R&60&53\\
2009 Okt 19&20:18\,--\,22:30&R&60&111\\
2009 Okt 20&20:05\,--\,23:30&R&60&178\\
2009 Okt 30&23:09\,--\,04:50&R&60&243\\
2009 Nov 06&00:45\,--\,02:55&R&60&95\\
2009 Nov 14&01:02\,--\,04:33&R&60&91\\
2010 Feb 03&20:42\,--\,22:12&R&60&74\\
2010 Feb 16&18:41\,--\,22:53&R&60&206\\
2010 Feb 20&20:05\,--\,22:30&R&60&120\\
2010 Feb 24&19:27\,--\,22:13&R&60&140\\
2010 Mar 02&22:21\,--\,23:09&R&60&41\\
2010 Mar 03&20:37\,--\,22:00&R&60&60\\
2010 Mar 06&20:40\,--\,20:52&R&60&11\\
2010 Mar 07&18:13\,--\,19:34&R&60&58\\
2010 Mar 23&18:44\,--\,20:49&R&60&74\\
2010 Aug 21&01:20\,--\,03:10&R&90&72\\
2010 Sep 11&00:54\,--\,02:32&V/I&180/50&10/10\\
2010 Sep 13&23:21\,--\,02:29&R&90&111\\
2010 Sep 15&23:00\,--\,02:31&R&90&75\\
2010 Sep 17&22:21\,--\,03:55&R&90&198\\
2010 Sep 18&00:32\,--\,03:50&R&90&75\\
2010 Sep 20&21:59\,--\,04:11&R&90&221\\
2010 Sep 21&21:58\,--\,04:11&R&90&219\\
2010 Sep 22&21:59\,--\,03:53&R&90&210\\
2010 Sep 23&01:06\,--\,03:47&R&90&96\\
2010 Okt 05&22:09\,--\,23:39&R&90&54\\
2010 Okt 08&21:12\,--\,01:00&R&90&134\\
2010 Okt 09&01:55\,--\,03:21&V/I&180/50&20/20\\
2010 Okt 10&20:05\,--\,22:40&R&90&88\\
2010 Okt 11&21:06\,--\,21:52&R&90&28\\
2010 Okt 13&20:06\,--\,22:41&R&90&92\\
2010 Okt 17&20:31\,--\,03:14&R&90&239\\
2010 Okt 20&20:39\,--\,01:12&R&90&114\\
2010 Okt 21&20:31\,--\,04:24&R&90&280\\
2010 Okt 22&20:29\,--\,02:57&R&90&230\\
2010 Okt 28&19:21\,--\,03:02&R&90&222\\
2010 Nov 02&20:00\,--\,22:34&R&90&79\\
2010 Nov 09&19:05\,--\,21:37&R&90&88\\
\hline  
\multicolumn{5}{l}{   
N~\footnotesize{is the number of images.}}
\end{tabular}
\end{table}

To determine the absolute magnitude of the field objects we used those well-known stars in our field that are constant (from their light curves in all observed nights). Those stars are listed in Table \ref{table1}. The apparent magnitude m can be calculated by the following equation
\begin{equation}
\mathrm{m} = \mathrm{C} +  \mathrm{m_{instr}} - \mathrm{k\cdot z}\, , 
\end{equation}
i.e. for the R-band we get
\begin{equation}
\mathrm{R} = \mathrm{C_R} +  \mathrm{R_{instr}} - \mathrm{k_{R}\cdot z} 
\end{equation}

where R is the magnitude to be measured and $\mathrm{R_{instr}}$ refers to 
the measured instrumental magnitude with our camera.\\ 
k and z are the extinction coefficient and airmass, respectively, 
whereas $\mathrm{C_R}$ denotes the zero point of the detector in the R-band.
Long-term variations of color terms are not yet available for this instrument, but should also be negligible for short-term events like flares.
For the determination of the zero point we used the constant known bright stars (Table \ref{table1}) in our Pleiades field, so that all stars in the field have the same extinction coefficient (k) due to the same airmass (z). To calibrate our images we added the zero point value (mean value in Table 1) to the measured differential magnitude ($\mathrm{m}$ - $\mathrm{C}$ + $\mathrm{k\cdot z}$) of every single star in the field.
\section{Results}
Our flare star is one of two unresolved stars (Fig. 1 and 2, unresolved in most images) with similar brightnesses in JHK-bands. In the 2MASS image (Fig. \ref{2MASS}), the two stars are well resolved. All information about both components are gathered in Table \ref{table5}. The coordinates and the magnitudes in infrared are taken from the 2MASS catalog (Cutri et al. 2003). The B magnitude (B\,=\,18.270\,mag) is from the NOMAD catalog (Zacharias et al. 2005). Since we are unable to resolve the two components, the measured magnitudes in VRI-bands belong to both components together (Table \ref{table4}). In order to calculate the individual magnitudes for each component we first used the individual JHK magnitudes of every component to determine its spectral type taking into account the optical extinction ($\mathrm{A_{V}}$) of the Pleiades of 0.2
mag (van Leeuwen 2009), which was used to calculate the extinction in other bands using Table 3 in Rieke and Lebofsky (1985). Then, by using the spectral type, the JHK magnitudes and Table A5 in Kenyon and Hartmann (1995) we determined the BVRI magnitudes of each component individually, which can be used to calculate the flux ratio of both
components in BVRI-bands. Finally, we determine the individual magnitudes of each component using the flux ratio and the measured magnitudes of both components together. The subtracted extinction is added again to all magnitudes, because we calculate the final spectral type and extinction by using the BVRIJHK magnitudes and Table 3 in Rieke and Lebofsky (1985) and Table A5 in Kenyon and Hartmann (1995). The final spectral type of both stars (Table 5) is M1--2 and the derived extinction is in agreement with that of the Pleiades. The effective temperature can be easily determined using Table A5 in Kenyon and Hartmann (1995) since the spectral type is known. Mass and radius are taken from Schmidt-Kaler (1982) assuming that both stars are main sequence stars.

\begin{figure*}
\centering
\includegraphics[width=13cm,angle=270]{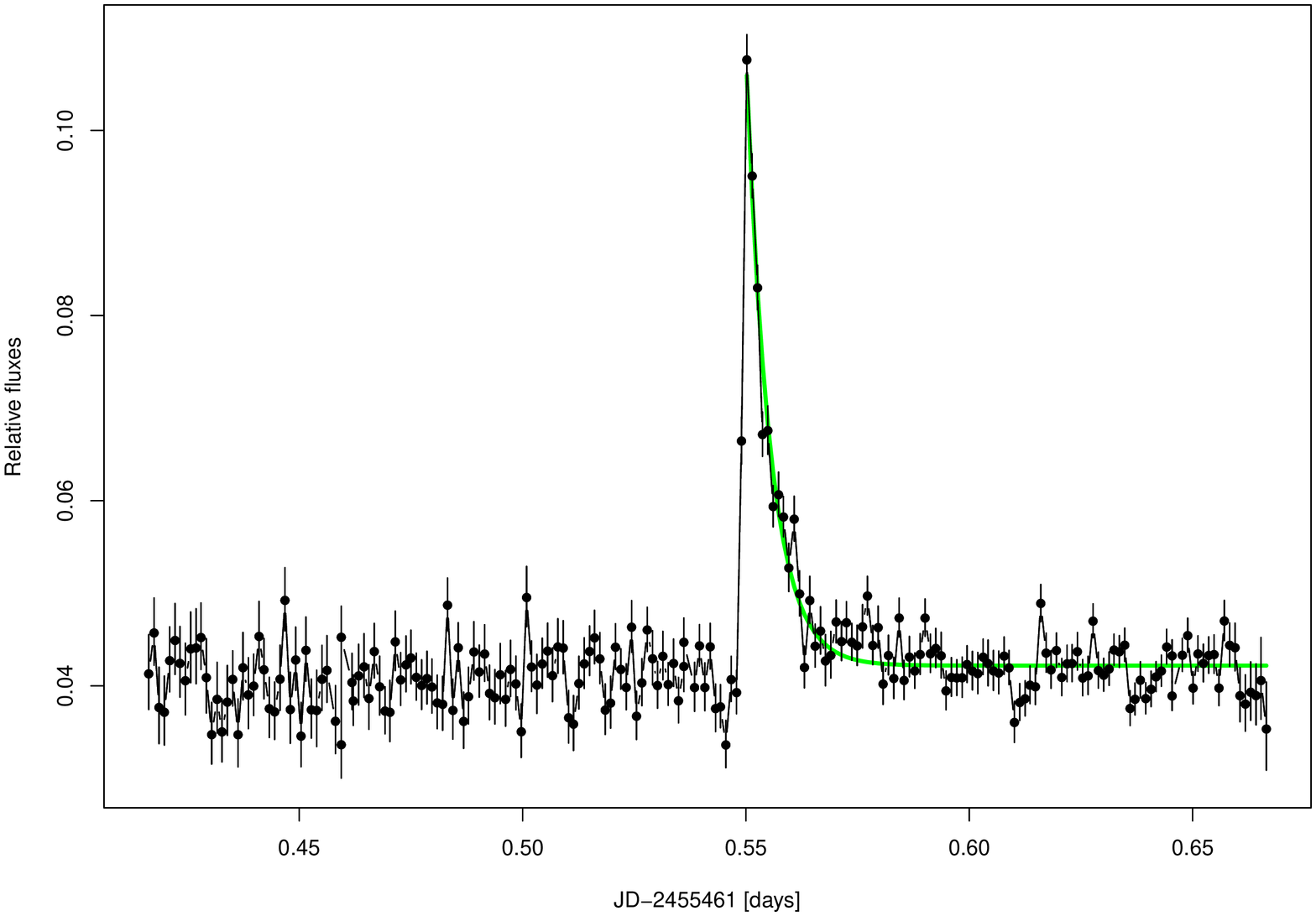}
\caption{Light curve with the flare observed in the night 2010 September 22 with the STK camera in R-band with an exposure time of 90 s. 
We plot relative flux, i.e. flare flux relative to the synthetic comparison star, which has 13.338 mag.
The flare began at 1:07(3)\,UT (JD 2455461.536), reached its maximum with an amplitude $\geq$ 1.08\,mag at 01:11(1.5)\,UT 
(JD 2455461.544), and ended at 02:03(4)\,UT (JD 2455461.581) with a total duration 
of 56\,$\pm$\,4\,min. (see Table \ref{table5}). The full line is the best fit to the flare decay, see text. 
The significance for the detection of a short small negative flare at JD 2455461.545 is only 2.6\,$\sigma$.}
\label{flare3}
\end{figure*}

\begin{table}
\caption{Parameters of the flare derived from its light curve in the night 2010 September 22 (JD 2455461).}
\label{table3}
\begin{tabular}{lll}\hline
Parameter & Value & JD-2455461 \\
\hline
negative flare [h:m]          & 01:05(2) UT   & 0.545 \\ 
beginning of flare [h:m]      & 01:08(3) UT   & 0.5475 \\
duration negative flare [min] & 8\,$\pm$2     & \\
maximum of flare [h:m]        & 01:12(1.5) UT & 0.550 \\ 
limit of increase [min]       & $\leq$ 4      & \\
duration of decrease [h:m]    & 00:59(2.5)    & \\
end of flare [h:m]            & 02:11(2) UT   & 0.5913 \\ 
total time [min]              & 63\,$\pm$4    & \\
amplitude [mag]               & $\geq$1.08    & \\ \hline
\end{tabular}
\end{table}

\begin{table}
\caption{Measured magnitudes of the flare star by using our STK camera (both components together unresolved), 2010 October 21.} 
\label{table4}
\begin{tabular}{ccc}\hline
 V&R&I\\
$\mathrm{[mag]}$&[mag]&[mag]\\
\hline
17.302\,$\pm$\,0.290&16.968\,$\pm$\,0.084&14.844\,$\pm$\,0.041\\
\hline
\end{tabular}
\end{table}

\begin{table}
\caption{\footnotesize{Characteristics of our flare star: VRI from our GSH data (split correctly among the two stars, see text), the B magnitude is taken from the NOMAD catalog (Zacharias et al. 2005), position, name, J, H, and Ks from 2MASS, spectral types and extinction from BVRIJHK (see text), temperature, mass, and radius for main sequence stars of those spectral types.}}
\label{table5}
\begin{tabular}{lcc}\hline
Parameter&star 1&star 2\\
\hline
$\alpha$ [h m s]&03 40 06.657&03 40 06.857\\
$\delta$ [$^\circ$ $'$ $''$]&+25 14 27.65&+25 14 25.63\\
2MASS-name&03400665+2514276&03400685+2514256\\
B [mag]&18.766\,$\pm$\,0.032&19.358\,$\pm$\,0.063\\
V [mag]&17.817\,$\pm$\,0.291&18.359\,$\pm$\,0.296\\
R\,\,[mag]&17.533\,$\pm$\,0.089&17.946\,$\pm$\,0.105\\
I\,~~[mag]&15.523\,$\pm$\,0.052&15.674\,$\pm$\,0.075\\
J\,~~[mag]&14.368\,$\pm$\,0.072&14.437\,$\pm$\,0.032\\
H\,~[mag]&13.764\,$\pm$\,0.088&13.833\,$\pm$\,0.050\\
K\,~[mag]&13.536\,$\pm$\,0.074&13.563\,$\pm$\,0.038\\
Sptype &M1V (M0...M2)&M2V (M1...M4)\\
$\mathrm{A_{V}}$ [mag]&0.231\,$\pm$\,0.024&0.266\,$\pm$\,0.020\\
$\mathrm{T_{eff}}$ [K]&3720 (3850...3580)&3580 (3720...3370)\\
M [$\mathrm{M_{\odot}}$]&0.45 (0.51...0.40)&0.40 (0.51...0.27)\\
R \,[$\mathrm{R_{\odot}}$]&0.55 (0.60...0.50)&0.50 (0.55...0.34)\\
\hline
\end{tabular}
\end{table}

\begin{figure*}
\centering
\includegraphics[width=9cm,height=15cm,angle=90]{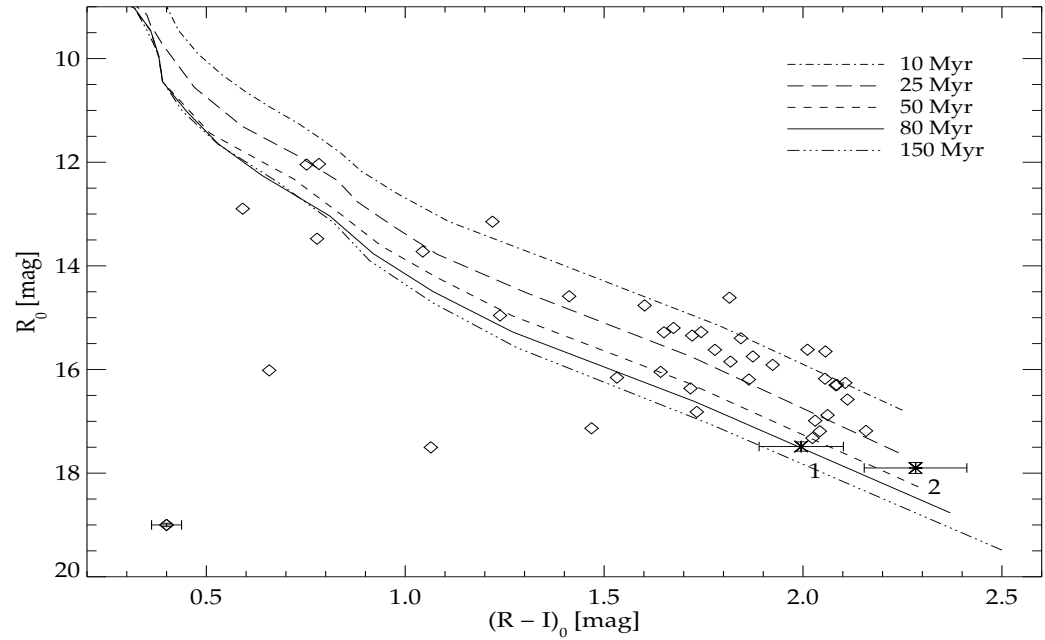}
\caption{The color-magnitude diagram (corrected for extinction) for The two data points well below the main sequence are probably
due to variability and non-simultaneous observations in R and I. Pleiades members in our field (diamonds) and the two 2MASS stars (asterisks) in our field, one of which has shown the flare. Five theoretical isochrones represent five ages (10, 25, 50, 80 and 150 Myr) and are taken from Siess et al. (2000) after adjusting their magnitudes to the distance of the Pleiades stars. Two data points well below the main sequence are probably
due to variability and non-simultaneous observations in R and I. The point in the lower left shows the mean error bar for our data for the Pleiades members.}
\label{Flare3}
\end{figure*}

\begin{figure*}
\centering
\includegraphics[width=9cm,height=15cm,angle=90]{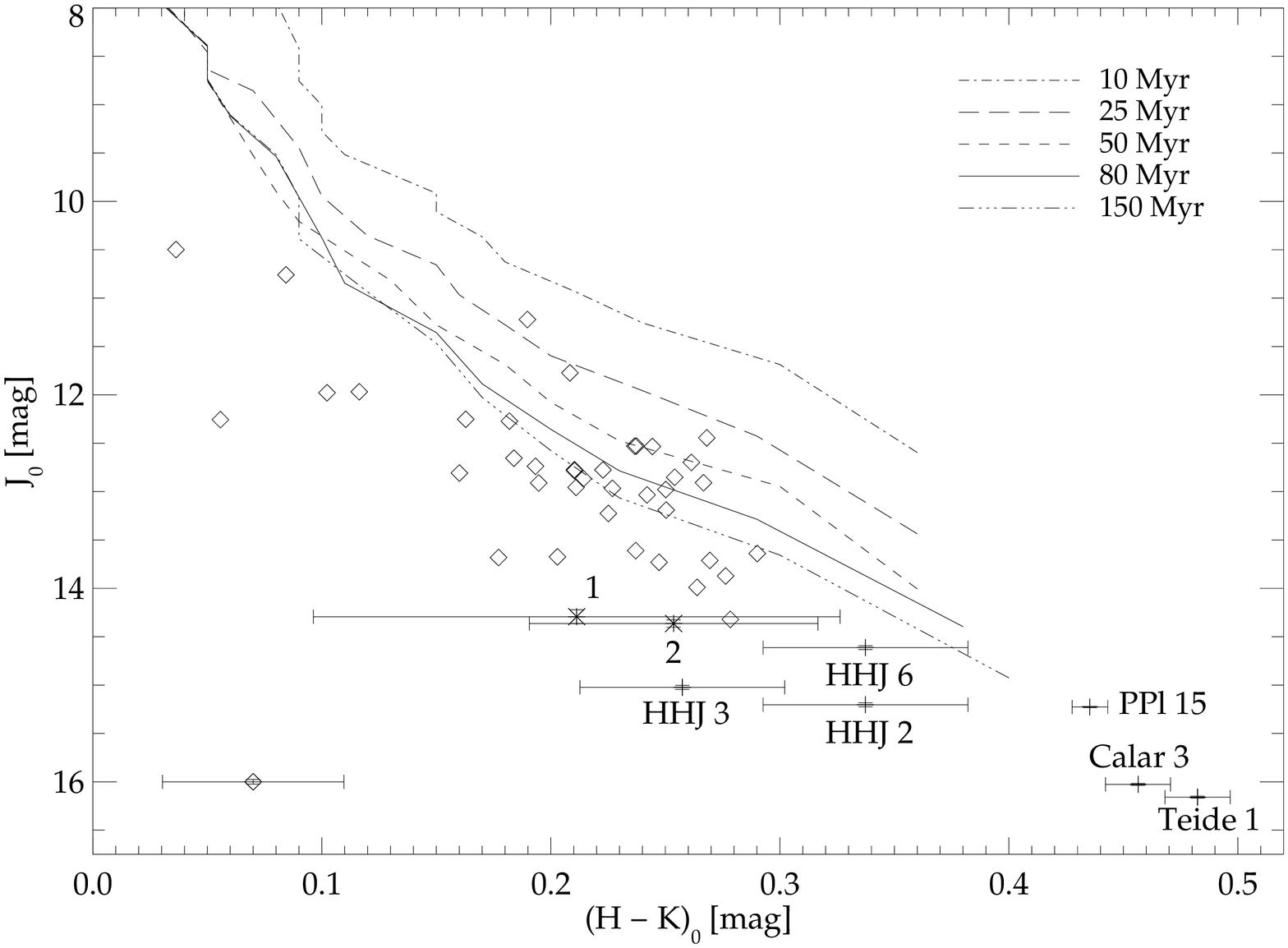}
\caption{The color-magnitude diagram (corrected for extinction) in the infrared for Pleiades members in our field (diamonds), and our two 2MASS stars (asterisks), one of which has shown the flare. Plusses refer to three low-mass stars (HHJ 2, 3 and 6), as well as PPl 15 and Teide 1 and Calar 3 in the Pleiades cluster (see text). The $JHK$ magnitudes of the Pleiades members are taken from the 2MASS catalog (Cutri et al. 2003), for HHJ 2, HHJ 6, PPl 15, Teide 1, Calar 3 and the $J$ magnitude of HHJ 3 from Zapatero Osorio et al. (1997), $HK$ magnitudes of HHJ 3 come from Steele et al. (1993). Isochrones are as in Fig. 5. The point in the lower left shows the mean error bar for the Pleiades members which are plotted here and detected by us in the optical.}
\label{Flare4}
\end{figure*}

On the night 2010 September 22, a flare was seen in the combined light curve of these two stars, whereas they show no variability but only a constant signal in all other observed nights. The light curve is typical for flares: Constant light before and after the flare, a short decrease in brightness (negative flare) before the sharp increase, then an exponential decay.\\
In Fig. \ref{flare3} we present the observed light curve of the flare, and its parameters indicated in Fig. \ref{flare3} are given as values in Table \ref{table3}.
We have first fitted relative fluxes (flare flux relative to the 
comparison star) of the flare decay part using the formula
\begin{equation}
Flux =  Const + A \cdot exp(B \cdot t)
\end{equation}
The flux, the constant, and the factor A are unitless, because we fit a relative flux,
i.e. the flux of the flare star relative to the synthetic comparison star; 
time $t$ is in days, constant B has then the unit 1/day.
With this non-linear model we obtained a very good fit with parameters: \\
A\,=\,0.0637\,$\pm$\,0.0022,  \\
B\,=\,-189.64\,$\pm$\,10.86 1/day,   \\
Const\,=\,0.04219\,$\pm$\,0.00031.  \\
In order to determine the flare end time, in addition to the above mentioned model,
we considered also purely constant as an alternative one. Thus, we compared
two models via F-statistics, i.e. the null hypothesis that the first model 
(exponential decay of the flare)
does not provide a significantly better fit than second one (purely constant).
We required the false rejection probability of the null hypothesis 
to lie below 0.05, the usual value.
We applied this approach subsequently to all data subsets starting from the
flare maximum up to the end of the observations.
The purely constant model can be accepted (rejecting the flare model) 
from the time JD-2455461 = 0.5913 to the end of observations, i.e. the flare decay time 
lasted 1 hour, which is typical for flares in the Pleiades cluster;
from that time on, the probability for the emission to be constant again
(after the flare) is always $\ge 0.95$. \\
Within the total exposure time of 132 hours we observed one flare with an amplitude of about 1.08\,mag in the R-band. Haro et al. (1982) have observed in the Pleiades a total of 1800 flares (in the blue wavelength range with an average amplitude of about 2.5\,mag) in 519 stars during a total exposure time of 3500 hours. i.e. one flare on average every 972 hours.\\ We have observed in our field 11-12 Pleiades flare stars (11 known and one probably new flare star of the Pleiades). Thus on average every (972 / 12 =) 81 hours a flare is expected - for a total exposure time of 132 hours thus 1.6 flares. This is not inconsistent with the actual observed flare.\\ 
None of the two stars (one of which has flared) is mentioned in the General Catalog of Variable Stars (GCVS) (Samus et al. 2009) as flare stars.\\
The proper motion of the second star 2MASS J03400685+\\2514256 is $\mu_{\alpha}\cdot\cos\delta$ = 43.9 $\pm$ 3.8 $\mathrm{mas}\cdot (\mathrm{yr})^{-1}$ and  $\mu_{\delta}$ = -- 44.3 $\pm$ 3.8 $\mathrm{mas}\cdot (\mathrm{yr})^{-1}$ (PPMXL catalog (Roeser et al. 2010)), which is inconsistent with the mean of that of the Pleiades members $\mu_{\alpha}\cdot\cos\delta$ = 20.10 $\pm$ 0.28 $\mathrm{mas}\cdot (\mathrm{yr})^{-1}$ and $\mu_{\delta}$ = -45.39 $\pm$ 0.27 $\mathrm{mas}\cdot (\mathrm{yr})^{-1}$ (van Leeuwen 2009 ).\\
Given the location and proper motion of this star (second star) with respect to the Pleiades, it also cannot be an ejected Pleiades member.\\
For the first star, 2MASS J03400665+2514276, there are no proper motion measurements available (NOMAD catalog, Zacharias et al. 2005).
However, it might be that the two stars were also unresolved in the observations used to determine the proper motion of
the second star, so that this proper motion may not be reliable.\\
The positions of the possible flare stars in the color magnitude diagram (Fig. \ref{Flare3} and \ref{Flare4}) are consistent with many known Pleiades members in our field as well as with low-mass Pleiades members, which are presented with three of the least massive stars (HHJ 2, 3, 6) in the Pleiades cluster (Hambly et al. 1993). They are separated from Pleiades BDs like Calar 3 and Teide 1 (Rebolo et al. 1996) by PPl 15 (Stauffer et al. 1994) which defines with its mass $M$\,=\,0.08~$M_\odot$ (Basri et al. 1996) the borderline between low-mass stars and BDs in the Pleiades cluster. \\Both stars are consistent with the age, the extinction and the distance of the Pleiades members.\\ We note that the Siess et al. (2000) isochrones in Fig. 6 (optical) seem to lie somewhat too low compared to our stars, very low-mass stars and brown dwarfs plotted (with literature data), while the Siess et al. (2000) isochrones in Fig. 7 (infrared) seem to lie too high, this is probably due to the models for very low masses.\\
In order to confirm or reject the first 2MASS star, whose proper motion is not yet known, as true Pleiades member, we plan to measure its proper motion with a new epoch IR image with high spatial resolution and to take a spectrum to confirm its spectral type. Since it is an early M-type flare star, it is probably a Pleiades member.

\textit{Acknowledgements}. 
This research has made use of the SIMBAD database, operated at CDS, Strasbourg, France, of data products from the Two Micron All Sky Survey, which is a joint project of the University of Massachusetts and the Infrared Processing and Analysis Center/California Institute of Technology, funded by the National Aeronautics and Space Administration, of the VizieR catalogue access tool, CDS, Strasbourg, France, and of the NASA/ IPAC Infrared Science Archive, which is operated by the Jet Propulsion Laboratory, California Institute of Technology, under contract with the National Aeronautics and Space Administration. For data reduction the ESO-MIDAS version 08SEPpl1 was used. Many thanks to the Observations group at the AIU Jena for observing the Pleiades field in the other nights. 
We would like to thank N. Vogt and R. Meinel for good suggestions.
We would also like to thank Martin Seeliger, Manfred Kitze, Donna Keeley, and Ina H\"ausler for also observing the Pleiades field in GSH. 
For the technical support we would like to thank Frank Gie\ss ler and J\"urgen Weiprecht for their help. 
RN would like to acknowledge financial support from the Thuringian
government (B 515-07010) for the STK CCD camera used in this project.
M. Moualla would like to thank Tishreen University in Syria-Lattakia for the 
scholarship and the DFG for their financial support in project NE 515 / 30-1. 
AB would like to thank DFG for support in NE 515 / 32-1.
CM would like to thank DFG for support in SCHR 665 / 7-1.
TR would like to thank DFG for support in project NE 515 / 36-1.
RN, RE, SR, and CA would like to thank DFG for support in the Priority Programme SPP 1385
on the {\em First ten Million years of the Solar System} in projects NE 515 / 34-1,
NE 515 / 33-1, and NE 515 / 35-1.
CG, TOBS, LT, TE, and TR would like to thank DFG for support in NE 515 / 30-1.
RN, VVH, MMH, LT, and JS would like to thank DFG for support from the SFB-TR 7.
NT would like to thank the Carl-Zeiss-Foundation for a scholarship.
SF thanks the State of Thuringia for a scholarship.
GM and TP would like to thank the European Union in the Framework Programme FP6
Marie Curie Transfer of Knowledge project MTKD-CT-2006-042514 for support.
VVH and LT would like to thank the DFG for financial support in project SFB TR 7.


\begin{thebibliography}{}

\bibitem[Ambartsumyan et al. (1970)]{1970Afz.....6....7A} Ambartsumyan, V.A., Mirzoyan, L.V., Parsamyan, E.S., Chavushyan, O.S., 
Erastova, L.K., 1970: Astrofizika 6, 7
\bibitem[[Ambartsumyan et al. (1971)]{1971Afz.....7..319A} Ambartsumyan, V.A., Mirzoyan, L.V., Parsamyan, E.S., Chavushyan, O.S., 
Erastova, L.K., 1971: Astrofizika 7, 319
\bibitem[Basri et al. 1996]{1996ApJ...458..600B} Basri, G., Marcy, G.W., Graham, J.R,. 1996: ApJ 458, 600  
\bibitem{} Bertin, E., 1997: {\it SExtractor v2.5}, Institut d`Astrophysique, Observatoire de Paris
\bibitem[Broeg et al. 2005] {2005AN....326..134B} Broeg, C., Fern{\'a}ndez, M., Neuh\"auser, R., 2005: AN 306, 134
\bibitem[Cutri et al. 2003]{2003tmc..book.....C} Cutri, R.M., Skrutskie, M.F., van Dyk, S., Beichman, C.A., et al., 2003: 
The 2MASS All-Sky Point Source Catalog, NASA/IPAC Infrared Science Archive Book
\bibitem[Eisenbeiss et al. 2009]{2009AN....330..439E} Eisenbeiss, T., Moualla, M., Mugrauer, M., Schmidt, T.O.B., Raetz, S., 
Neuh{\"a}user, R., et al., 2009: AN 330, 439
\bibitem[Hambly et al. 1996]{1993A&AS..100..607H} Hambly, N.C., Hawkins, M.R.S., Jameson, R.F., 1993: A\&AS 100, 607
\bibitem[Haro et al. 1982]{1982BITon...3....3H}Haro, G., Chavira, E., Gonzalez, G., 1982: Boletin del Instituto de Tonantzintla 3, 3
\bibitem[Hertzsprung, 1924]{1924BAN.....2...87H} Hertzsprung, E., 1924: BAN 2, 87
\bibitem[Johnson and Mitchell 1958]{1958ApJ...128...31J} Johnson, H.L., Mitchell, R.I., 1958: ApJ 128, 31
\bibitem[Kenyon and Hartmann 1995]{1995ApJS..101..117K} Kenyon, S.J., Hartmann, L., 1995: ApJS 101, 117
\bibitem[]{1995Ap.....38..276M} Mirzoyan, L.V., Hambaryan, V.V., Garibjanian, A.T., Mirzoyan, A.L., 1995: Astrophysics 38, 276
\bibitem{}Moualla, M., 2011: PhD thesis, University Jena
\bibitem[Mugrauer 2009]{2009AN....330..419M} Mugrauer, M., 2009: AN 330, 419
\bibitem[Mugrauer \& Berthold]{2010AN....331..449M} Mugrauer, M., Berthold, T., 2010: AN 331, 449
\bibitem[Rebolo et al. 1995]{1995Natur.377..129R}Rebolo, R., Zapatero Osorio, M.R., Mart{\'{\i}}n, E.L., 1995: Nature 377, 129
\bibitem[Rebolo et al. 1996]{1996ApJ...469L..53R}Rebolo, R., Martin, E.L., Basri, G., Marcy, G.W., Zapatero-Osorio, M.R., 1996: ApJ 469, L53
\bibitem[Rieke \& Lebofsky 1985]{1985ApJ...288..618R} Rieke, G.H., Lebofsky, M.J., 1985: ApJ 288, 618
\bibitem[Roeser et al. 2010]{2010AJ....139.2440} Roeser, S., Demleitner, M., Schilbach, E., 2010: AJ 139, 2440
\bibitem[Samus 2009]{2009yCat....102025S} Samus, N.N., Durlevich, O.V., et al., 2009: General Catalogue of Variable Stars
\bibitem[Schmidt-Kaler, 1982]{} Schmidt-Kaler, 1982, in Landolt-B\"ornstein
\bibitem[Schwarz \& Becklin 2005]{2005AJ....130.2352S} Schwartz, M.J., Becklin, E.E., 2005: AJ 130, 2352
\bibitem[Siess 2000]{2000A&A...358..593S} Siess, L., Dufour, E., Forestini, M., 2000: A\&A 358, 593
\bibitem[Skrutskie 2006]{2006AJ....131.1163S} Skrutskie, M.F., Cutri, R.M., Stiening, R., Weinberg, M.D., et al., 2006: AJ 131, 1163
\bibitem[Stauffer 1991]{1991AJ....101..980S} Stauffer, J., Klemola, A., Prosser, C., Probst, R., 1991: AJ 101, 980
\bibitem[Stauffer et al. 1994]{1994AJ....108..155S} Stauffer, J.R., Hamilton, D., Probst, R.G., 1994: AJ 108, 155
\bibitem[Steele et al. 1993]{1993MNRAS.263..647S} Steele, I.A., Jameson, R.F., Hambly, N.C., 1993: MNRAS 263, 647
\bibitem[Van Leeuwen 2009]{2009A&A...497..209V} van Leeuwen, F., 2009: A\&A 497, 209
\bibitem[Zacharias 2005]{2005yCat.1297....0Z} Zacharias, N., Monet, D.G., Levine, S.E., Urban, S.E., Gaume, R., Wycoff, G.L., 2005: NOMAD Catalog
\bibitem[M. R. Zapatero Osorio et al. 1997]{1997A&A...317..164Z} Zapatero Osorio, M.R., Rebolo, R., Martin, E.L., 1997: A\&A 317, 164
\end{thebibliography}
\end{document}